\documentclass[sigconf]{acmart} %
\usepackage{booktabs} 

\usepackage{xcolor}
\usepackage[ruled, noline, linesnumbered]{algorithm2e}
\usepackage{xcolor}

\definecolor{lightgreen}{RGB}{225,250,229}
\definecolor[named]{Green}{cmyk}{0.8,0,1,0.19}
\def\HighLight{\leavevmode\rlap{\hbox to \hsize{\color{lightgreen}\leaders\hrule height 0.8\baselineskip depth .5ex\hfill}}}

\usepackage{listings}
\usepackage{pifont}

\let\othelstnumber=\thelstnumber
\def\createlinenumber#1#2{
    \edef\thelstnumber{%
        \unexpanded{%
            \ifnum#1=\value{lstnumber}\relax
              #2%
            \else}%
        \expandafter\unexpanded\expandafter{\thelstnumber\othelstnumber\fi}%
    }
    \ifx\othelstnumber=\relax\else
      \let\othelstnumber\relax
    \fi
}

\usepackage{wrapfig}

\usepackage{tikz}

\usetikzlibrary{arrows}

\newcommand{\shadowjpf}[0]{\textsc{$Shadow_{JPF}$}}
\newcommand{\shadowklee}[0]{ \textsc{$Shadow_{KLEE}$}}

\usepackage{balance}
\usepackage{hyperref}

\begin{document}

\setcopyright{none}

\acmDOI{} 

\acmISBN{}

\acmYear{}

\title{Shadow Symbolic Execution with Java PathFinder}

\author{Yannic Noller}
\affiliation{
  \institution{Humboldt University of Berlin}
  \city{Berlin}
  \country{Germany}
}
\email{yannic.noller@informatik.hu-berlin.de}

\author{Hoang Lam Nguyen}
\affiliation{
  \institution{Humboldt University of Berlin}
  \city{Berlin}
  \country{Germany}
}
\email{nguyenhx@informatik.hu-berlin.de}

\author{Minxing Tang}
\affiliation{
  \institution{Humboldt University of Berlin}
  \city{Berlin}
  \country{Germany}
}
\email{tangminx@informatik.hu-berlin.de}

\author{Timo Kehrer}
\affiliation{
  \institution{Humboldt University of Berlin}
  \city{Berlin}
  \country{Germany}
}
\email{timo.kehrer@informatik.hu-berlin.de}

\begin{abstract}
Regression testing ensures that a software system when it evolves still performs correctly and that the changes introduce no unintended side-effects.
However, the creation of regression test cases that show divergent behavior needs a lot of effort.
A solution is the idea of \textsl{shadow symbolic execution}, originally implemented based on KLEE for programs written in C, which takes a unified version of the old and the new program and performs symbolic execution guided by concrete values to explore the changed behavior.
In this work, we apply the idea of shadow symbolic execution to Java programs and, hence, provide an extension of the Java PathFinder (JPF) project to perform shadow symbolic execution on Java bytecode.
The extension has been applied on several subjects from the JPF test classes where it successfully generated test inputs that expose divergences relevant for regression testing.
\end{abstract}


\maketitle

\section{Introduction}
One of the distinctive properties of real-world software is that it \textsl{evolves}, since it has to be adapted to its continuously changing environment.
Software changes, usually referred to as \textsl{patches}, typically fix incorrect behavior or introduce new functionality. However, it is also known that these patches are prone to introduce new errors \cite{Gu2010, Yin2011}, which is why users are often hesitant to update to the latest version.

To prevent this problem, \textsl{regression testing} is performed on the modified program version in order to provide confidence that the newly introduced software changes behave as expected and have no unintended side-effects. Since this is an expensive process, it is important to select the appropriate test cases. For instance, several regression testing techniques \cite{Harrold2001,Graves2001} select and run a subset of the test cases from the program's existing test suite or automatically generate test cases with high coverage of the changed code \cite{Marinescu2013}. However, even if the selected test cases achieve full statement or full branch coverage of the patch code, they do not necessarily exercise all new behaviors introduced by the patch.

To give an illustration, consider a patch that only changes the conditional statement \texttt{if(x > 5)} to \texttt{if(x > 10)}. The two test cases \texttt{x=0} and \texttt{x=15} cover both sides of the branch, but the execution of these inputs is completely unaffected by the patch since they result in the same branching behavior in both program versions. On the other hand, if \texttt{x} is between 6 and 10 (inclusive), the two program versions exhibit divergent behavior as they take different sides of the branch.

Recently, Palikareva et al. \cite{Palikareva2016} have introduced a dynamic symbolic execution-based technique, which they refer to as \textsl{shadow symbolic execution}. Their technique is designed to generate test inputs that cover new program behaviors introduced by a patch. Shadow symbolic execution works by executing both the old (bug\-gy) and new (patched) version in the same symbolic execution instance, with the old version \textsl{shadowing} the new one. Therefore, it is necessary to manually \textit{merge} both programs into a change-annotated, unified version. Based on such a unified version, the technique detects divergences along the execution path of an input that exercises the patch. Their tool \textsc{Shadow}, which we refer to as \shadowklee, is implemented on top of the KLEE symbolic execution engine \cite{Cadar2008KLEE}.

Our novel implementation \shadowjpf, as an extension of the Java PathFinder (JPF) \cite{Visser2003}, applies the idea of shadow symbolic execution to Java bytecode and, hence, allows to detect divergences in Java programs that expose new program behavior. The application of our extension on various subjects from the JPF test classes evaluate its test case generation capabilities.
\section{Shadow Symbolic Execution}\label{sec:shadow-symbolic-execution}

Shadow symbolic execution \cite{Palikareva2016} aims at generating test inputs that cover the new program behaviors introduced by a patch. Their approach takes as input the buggy and the patched version (say $old$ and $new$, respectively) and assumes an existing test suite.

\createlinenumber{9}{9+}
\lstset{numbers=left,xleftmargin=2em,frame=lrtb, framexleftmargin=1em}
\begin{minipage}{\linewidth}
\begin{lstlisting}[language=Java,basicstyle=\scriptsize\ttfamily, keywordstyle=\bf,numbers=left,frame=none,tabsize=3,firstnumber=1,xleftmargin=5em, captionpos=b, caption={Toy example to show the approach of shadow symbolic execution.}, label={lst:example}]{Name}
int foo(int x){
  int y;
  if(x < 0){
     y = -x;
  }
  else{
     y = 2 * x;
  }
  <@\color{Green}{y = -y;}@>
  if(y > 1){
      return 0;
  } else {
      if(y == 1 || y <= -2){
          assert(false);
      }
  }
  return 1;
}
\end{lstlisting}
\end{minipage}
\createlinenumber{9}{9}

\begin{figure*}[!t]
		\centering
		\begin{tikzpicture}[scale=0.7, every node/.style={transform shape}]
		\tikzset{vertex/.style = {shape=circle,draw,minimum size=1.5em}}
		\tikzset{edge/.style = {->,> = latex'}}
		
		\node[draw,rectangle,align=center] at (-1.7,-2) (n1) {
			...};	
		
		\node[draw,rectangle,align=center] at (-1.7,-3.5) (n2) {
			\footnotesize $[PC_{old}: (X < 0)]$ \\
			\footnotesize $[PC_{new}: (X < 0)]$ \\ 
			{\footnotesize \color{Green}SAT $[x<0]$} \\
			\footnotesize $10_{old}: -X > 1$ ? \\
			\footnotesize $10_{new}: X > 1$ ?};	
			
		\node[draw,rectangle,align=center] at (-9.1,-7.0) (n3) {
			\footnotesize $[PC_{old}: (X < 0) \land (-X > 1)]$ \\
			\footnotesize $[PC_{new}: (X < 0) \land (X > 1)]$ \\ 
			{\footnotesize \color{red}UNSAT}}; 
			
		\node[draw,rectangle,align=center] at (-4.0,-7.0) (n4) {
			\footnotesize $[PC_{old}: (X < 0) \land (-X \leq 1)]$ \\
			\footnotesize $[PC_{new}: (X < 0) \land (X \leq 1)]$ \\ 
			{\footnotesize \color{Green}SAT $[x = -1]$} \\
			\footnotesize $13_{old}: (-X == 1 || -X <= -2) $ ? \\
			\footnotesize $13_{new}: (X == 1 || X <= -2) $ ?};

		\node[draw,rectangle,align=center] at (-4.0,-8.5) (n5) {
			...};	
			
		\node[draw,rectangle,align=center] at (1.0,-7.0) (n7) {
			\footnotesize $[PC_{old}: (X < 0) \land (-X \leq 1)]$ \\
			\footnotesize $[PC_{new}: (X < 0) \land (X > 1)]$ \\ 
			{\footnotesize \color{red}UNSAT}}; 

		\node[draw,rectangle,align=center] at (6.0,-7.0) (n8) {
			\footnotesize $[PC_{old}: (X < 0) \land (-X > 1)]$ \\
			\footnotesize $[PC_{new}: (X < 0) \land (X \leq 1)]$ \\ 
			{\footnotesize \color{Green}SAT $[x \leq -2]$} \\
			\footnotesize $13_{new}: (X == 1 || x <= -2) $ ?};
			
		\node[draw,rectangle,align=center] at (6.0,-8.5) (n9) {
			...};	
		
		\draw[edge] (n1.south) -- (n2.north) node [midway,left]  {};
		\draw[edge] (n2.west) -- (n3.north) node [midway,left] {\footnotesize$same_{true}$\phantom{mm}};	
		\draw[edge] (n2.south) -- (n4.north) node [midway,left] {\footnotesize$same_{false}$\phantom{m}};		
		\draw[edge] (n4.south) -- (n5.north) node [midway,left]  {};
		\draw[edge] (n2.south) -- (n7.north) node [midway,right] {\footnotesize$diff_{true}$\phantom{m}};
		\draw[edge] (n2.east) -- (n8.north) node [midway,right] {\footnotesize\phantom{m}$diff_{false}$};
		\draw[edge] (n8.south) -- (n9.north) node [midway,left]  {};
		
		\end{tikzpicture}
\hspace{2em}
\caption{Partial four-way forking symbolic execution tree for the combined execution of the old and the new version of the program in Listing \ref{lst:example} for the test input $x=-1$. Each node represents a state in the symbolic search space, where each state holds the combined information of the old and the new symbolic execution.}
\label{fig:executiontree}
\end{figure*}
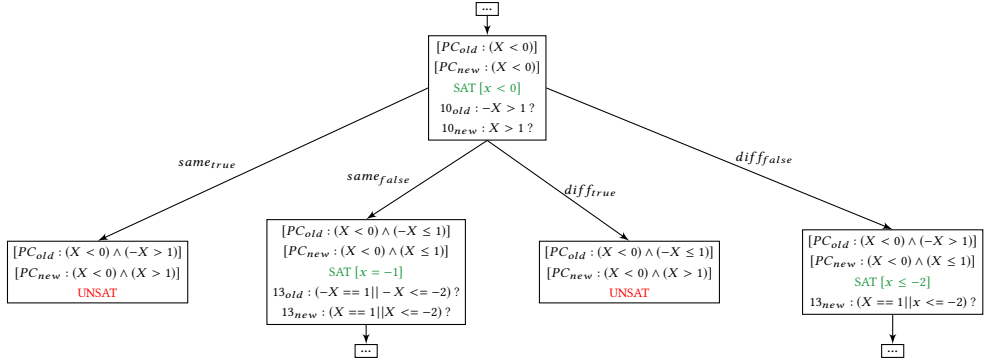

To give an illustration, consider the patch for the method \texttt{foo()} in Listing \ref{lst:example}.
There is an additional assignment in line 9 for the variable $y$ that negates it to $-y$.
This patch fixes the assertion error (line 14) for $x = -1$, but it introduces a new assertion error for, e.g., $x = -2$. Since the approach aims at generating test cases for the different execution paths of the buggy and the patched version, the optimal result would be two test inputs: (i) one for the fixed path, and (ii) one for the path with the newly introduced assertion error. 

In order to execute both program versions in a single symbolic execution instance, Palikareva et al. \cite{Palikareva2016} follow an approach that unifies both program versions with \texttt{change()} annotations. The annotations resemble a function call with two arguments, where the first argument represents the code expression from the old version and the second argument the expression from the new version.
The unifying process is performed manually. In our example, the change of line 9 is annotated as \texttt{y=change(y,-y)}.

Afterwards, the dynamic symbolic execution is performed in two steps: (i) the concolic phase and (ii) the bounded symbolic execution (BSE) phase. The concolic phase is initialized with the test cases that touch at least one patch statement. It collects the divergence points that are later used as starting points for the BSE phase.
All types of instructions are basically handled in the same way as in traditional symbolic execution, except for conditional statements, for which shadow symbolic execution forks execution into four different paths, where each path is specified by the side of the branch taken by the different program versions (cf. Figure \ref{fig:executiontree}). 
A $\mathit{same}$ path is the execution path where the new and the old program version take the same side of the conditional statement, i.e., if both versions follow the $true$ branch (aka $\mathit{same}_{true}$ path) or vice versa both versions follow the $false$ branch (aka $\mathit{same}_{false}$ path).
A $\mathit{diff}$ path is an execution path where the new program version takes an execution path which is different from that of the old version, i.e., if the new version follows the $true$ branch of a conditional statement while the old version follows the $false$ branch (aka $\mathit{diff}_{true}$ path) or vice versa (aka $\mathit{diff}_{false}$ path).
If the concrete executions of the two versions diverge at a conditional statement, the concolic phase will be stopped and the divergence point for the $\mathit{diff}_{x}$ path will be added to the BSE phase. If both program versions behave identically for the concrete input, shadow symbolic execution follows the $same_{x}$ path. However, if divergences are possible, for each feasible $\mathit{diff}$ path an input that exercises the divergent behavior will be generated and added to the BSE phase.

This strategy includes that for adding and removing straightline code, encapsulated by the change annotations \texttt{if(change(false,\\true))} and \texttt{if(change(true,false))}, shadow symbolic execution directly triggers a divergence point. 
This is a conservative approach, hence, an over-approximation of the $\mathit{diff}$ paths, since the added/deleted code may not propagate a change at a branching point and thus not lead to a $\mathit{diff}$ path.
As long as the concolic executions do not diverge, shadow symbolic execution continues executing both program versions until the end of the program in order to explore any additional divergences along the way. 
The BSE phase runs only on the new version starting from the collected conditional statements, i.e., symbolic execution on a fixed resource budget with a breadth-first exploration of the execution tree. 

At the end, the generated test inputs are used to execute both versions and the results are manually compared to classify them as expected divergences (such as an intended bug fix) or as regression bugs.
A solution is the idea of shadow symbolic execution, originally implemented based on KLEE for programs written in C.

As an illustration, Figure \ref{fig:executiontree} shows a partial four-way forking symbolic tree for the combined execution of the $old$ and the $new$ version of the program in Listing \ref{lst:example}.
Depending on the concrete input, shadow symbolic execution only explores a subset of the possible symbolic states. This has the benefit to limit the search space.
Suppose that a developer has written the test case $x = -1$, as this input caused an assertion error in the old program version. Note that this input also fully covers the changed statements.
At the first conditional statement in line 3, shadow symbolic execution simply follows the concrete execution since there was no change yet and, hence, the $\mathit{diff}$ paths are unsatisfiable at this point. After executing line 9, the variable $y$ is mapped to the symbolic expression $-X$ in the old program version and to $X$ in the new version with the concrete values now being $1$ and $-1$, respectively.
As a result, both concrete executions take the $false$ branch at the conditional statement in line 10, i.e. they follow the $\mathit{same}_{false}$ path. Shadow symbolic execution identifies additionally, that the $\mathit{diff}_{false}$ path is satisfiable (cf. Figure \ref{fig:executiontree}) and stores the divergence point for the second execution phase.
Continuing with the concrete executions this leads to the path that is only followed by the input $x=-1$. This execution will be stopped at line 13, where the concrete executions diverge: the old version takes the $true$ branch that leads to an assertion error and the new version takes the $false$ branch that returns $1$ and represents the bug fix. Shadow symbolic execution will report a $\mathit{diff}$ path for $x=-1$, which can be classified as an expected change.
Eventually, bounded symbolic execution will be started from the stored divergence point from line 10. With the given path condition only one path is left as feasible, which is $x \leq -2$. This $\mathit{diff}$ path leads to an assertion error and can be classified as a regression bug.
\section{Implementation}
\label{sec:implementation}
\shadowjpf~is implemented as an extension of the symbolic execution project of JPF, namely Symbolic PathFinder (SPF) \cite{Pasareanu2013}, and leverages its symbolic execution functionality in order to enable shadow symbolic execution of Java bytecode. Similar to SPF, the tool makes use of various extension mechanisms of JPF, such as attribute objects, choice generators and listeners. In fact, \shadowjpf~ \\
overrides the core extensions of SPF in order to specifically support shadow symbolic execution.
The tool is available at: \\
\url{https://github.com/hub-se/jpf-shadow}

\subsection{Efficiently sharing Symbolic States using DiffExpressions}
Since shadow symbolic execution runs both program versions (as a single unified program) in the same symbolic execution instance, it is important to maximize sharing between the symbolic states in order to keep memory consumption low.
Similar to the approach of Palikareva et al. \cite{Palikareva2016}, instead of maintaining two separate symbolic stores, \shadowjpf~constructs a \texttt{DiffExpression} whenever the tool encounters a \texttt{change()} annotation.
A \texttt{DiffExpression} basically stores the symbolic and shadow expression of a variable and is associated with it as a data attribute object in the same way as a regular symbolic expression in SPF.
Note that storing a \texttt{DiffExpression} is only necessary if the symbolic expression of a variable diverges between the two program versions.
As long as the symbolic expression of a variable is equal in both program versions, storing a single symbolic expression object (as provided by \textsc{SPF}) is sufficient.
Algorithm 1 illustrates how a \texttt{DiffExpression} is constructed whenever a \texttt{change()} method is invoked.

\IncMargin{1.5em}
\begin{algorithm}[h]
\SetKwData{Old}{old}
\SetKwData{New}{new}
\SetKwData{AttrOld}{attr\_old}
\SetKwData{AttrNew}{attr\_new}
\SetKwData{rShadow}{result\_shadow}
\SetKwData{rSymbc}{result\_symbc}
\SetKwFunction{DiffExpr}{DiffExpression}
\SetKwFunction{getShadow}{getShadow}
\SetKwFunction{getSymbc}{getSymbc}
\caption{Execute change(old,new) method}

\AttrOld $\gets$ attribute object of \Old \;
\AttrNew $\gets$ attribute object of \New \;

\uIf{\AttrOld instanceof \DiffExpr}{
\rShadow $\gets$ \AttrOld .\getShadow{}\; 
}
\uElse{
\rShadow $\gets$ \AttrOld
}

\uIf{\AttrNew instanceof \DiffExpr}{
\rSymbc $\gets$ \AttrNew .\getSymbc{}\; 
}
\uElse{
\rSymbc $\gets$ \AttrNew
}
\Return new \DiffExpr{\rShadow,\rSymbc}
\label{fig:diffExpression}
\end{algorithm}

\subsection{Extended Bytecode Implementation}
\textbf{Arithmetic bytecode}. Since a symbolic variable can be associated to either a symbolic expression or a \texttt{DiffExpression}, the arithmetic as well as the branching bytecode has to be extended in order to support both types of expressions. As an example, consider Algorithm 2 that describes how shadow symbolic execution of the \texttt{IADD} instruction is performed. The highlighted lines show the differences between the implementation of SPF and \shadowjpf. Similar to the implementation in SPF, it is first checked whether both operands are concrete, in which case the execution is delegated to the concrete super class. Otherwise, if at least one operand is symbolic, the result also becomes symbolic. The key idea is to determine the shadow and symbolic expression for both operands (line 9 to 14) and simply add the respective expressions to obtain the symbolic and shadow expression of the result (line 15 and 16). Note that the shadow and symbolic expression of a variable are equal if they have not diverged yet (line 13 and 14). For this reason, only if at least one of the operand attributes is a \texttt{DiffExpression}, the resulting attribute object also becomes a \texttt{DiffExpression} (line 17 to 20).

\begin{table}[h]
\resizebox{\columnwidth}{!}{%
\begin{tabular}{c | c | l}
Choice & Path & PC \\ \hline
1 & $\mathit{same}_{true}$ & $pc \land (\mathit{sym\_v1} = \mathit{sym\_v2}) \land (\mathit{shadow\_v1} = \mathit{shadow\_v2})$ \\
2 & $\mathit{same}_{false}$ & $pc \land (\mathit{sym\_v1} \neq \mathit{sym\_v2}) \land (\mathit{shadow\_v1} \neq \mathit{shadow\_v2})$ \\
3 & $\mathit{diff}_{true}$ & $pc \land (\mathit{sym\_v1} = \mathit{sym\_v2}) \land (\mathit{shadow\_v1} \neq \mathit{shadow\_v2})$ \\
4 & $\mathit{diff}_{false}$ & $pc \land (\mathit{sym\_v1} \neq \mathit{sym\_v2}) \land (\mathit{shadow\_v1} = \mathit{shadow\_v2})$ \\
5 & concrete & depends on the concrete input \\
\end{tabular}}
\caption{The five possible choices for each execution path and the corresponding path conditions for the \texttt{IF\_ICMPEQ} instruction. pc denotes the current path condition while sym\_v1/v2 and shadow\_v1/v2 represent the symbolic and shadow expressions of the operands to be compared, respectively.}
\label{fig:choices}
\end{table}

\noindent\textbf{Branching bytecode}. In SPF, the symbolic execution of a conditional statement involves setting a \texttt{ChoiceGenerator} with two choices, which represent the $\mathit{true}$ and $\mathit{false}$ sides of the branch.
Each choice is associated with the respective path condition, which is checked for satisfiability by a constraint solver.
Recall that shadow symbolic execution forks execution into four different paths, where each path is specified by the side of the branch taken by the two program versions (cf. Figure \ref{fig:executiontree}).
Therefore, it is necessary to create four choices, one for each possible path.
Figure \ref{fig:choices} gives an overview of the choices and the resulting path conditions for the \texttt{IF\_ICMPEQ} instruction that compares to variables for equality.
Originally, shad\-ow symbolic execution operates in two phases: (i) the concolic phase and (ii) the bounded symbolic execution (BSE) phase.
Instead of running a concolic phase, we added a fifth choice that determines the next execution path based on the concrete inputs.
To give an illustration, consider the modified branching statement \texttt{if(change(x>1, x<=5))} with the concrete input $x = 3$.
In order to determine the outcome in both program versions, we check the satisfiability of the constraints $(x>1 \land x=3)$ and $(x<=5 \land x=3)$ for the old and new program version, respectively.
Since both constraints are satisfiable, both versions take the $\mathit{true}$ path.
Note that in this case the concrete execution replaced the exploration of the $\mathit{same}_{true}$ path (choice 1).
As a result, if we only consider the choices 5, 4 and 3 (cf. Figure \ref{fig:choices}) at each conditional statement, \shadowjpf~can follow the concrete execution of both program versions until they diverge, while checking for possible $\mathit{diff}$ paths along the concrete execution path.
As soon as a $\mathit{diff}$ path is explored, only the choices 1 and 2 are considered (while ignoring the shadow expressions), effectively replacing the bounded symbolic execution phase. \\

\begin{algorithm}[h]
\SetKwData{firstOp}{op\_v1}
\SetKwData{secondOp}{op\_v2}
\SetKwData{opvi}{$\text{op\_v$_i$}$}
\SetKwData{symvi}{$\text{sym\_v$_i$}$}
\SetKwData{shadowvi}{$\text{shadow\_v$_i$}$}
\SetKwData{firstSym}{sym\_v1}
\SetKwData{firstShadow}{shadow\_v1}
\SetKwData{secondSym}{sym\_v2}
\SetKwData{secondShadow}{shadow\_v2}
\SetKwData{rShadow}{shadow\_r}
\SetKwData{rSymbc}{sym\_r}
\SetKwData{result}{result}
\SetKwData{stack}{stack}
\SetKwFunction{pop}{pop}
\SetKwFunction{push}{push}
\SetKwFunction{DiffExpr}{DiffExpression}
\SetKwFunction{getShadow}{getShadow}
\SetKwFunction{getSymbc}{getSymbc}
\SetKwFunction{execute}{execute}
\SetKwFunction{setAttr}{setAttributeObject}
\SetKwFunction{nextInsn}{nextInstruction}
\SetKwComment{Comment}{$\triangleright$\ }{}

\caption{Shadow symbolic execution of \texttt{IADD}}

\firstOp $\gets$ attribute object of first operand \;
\secondOp $\gets$ attribute object of second operand \;

\uIf{operands concrete}
{
	\Return super.\execute{} ;
}
\uElse{
	\stack .\pop{}\;
	\stack .\pop{}\;
	\stack .\push{0}\;
	\HighLight\uIf{\opvi \emph{$\left(i \in 1,2 \right)$} instanceof \DiffExpr } 
	{
		\HighLight\symvi $\gets$ \opvi .\getSymbc{} \;
		\HighLight\shadowvi $\gets$ \opvi .\getShadow{} \;
	}
	\uElse{
		\HighLight\symvi $\gets$ \opvi \;
		\HighLight\shadowvi $\gets$ \opvi \;
	}
	\rSymbc $\gets$ \firstSym $+$ \secondSym \;
	\HighLight\rShadow $\gets$ \firstShadow $+$ \secondShadow \;
	\HighLight\uIf{\firstOp or \secondOp instanceof \DiffExpr}
	{
	\HighLight	\result $\gets$ new \DiffExpr{\rShadow,\rSymbc}\;
	}
	\uElse{
		\result $\gets$ \rSymbc\;
	}
	\setAttr{\result}\;
	\Return \nextInsn{}\;
}

\label{alg:iadd}
\end{algorithm}
\vspace{-0.5cm}
\section{Evaluation}
We evaluated our implementation w.r.t (i) its correctness and (ii) its effectiveness of generating regression tests for Java programs. Therefore, we answer the following research questions:

\noindent\textbf{RQ.1}: Is our implementation consistent with the original implementation that was implemented for C programs? \\
\textbf{RQ.2}: Compared to pure SPF, can \shadowjpf~generate more test cases that are relevant for regression testing?

\subsection{Experimental Setup}
For the evaluation we used publicly available artifacts that JPF/ SPF can handle and made them usable for regression testing by generating multiple versions of them with the \textsc{Major} mutation framework \cite{Just2011}. We used the full generation setup as provided by the authors, without the operators that cannot be handled by SPF or which produced errors in our SPF experiments. As a first evaluation step, we selected the following software artifacts as our experimental subjects from the official SPF repository\footnote[1]{https://babelfish.arc.nasa.gov/hg/jpf/jpf-symbc}: BankAccount.deposit(), BankAccount.withdraw(), BankAccount.main() and generated in total 51 mutants. For all mutants we manually added the change annotations to generate the unified version. 
Since only the executable binaries of the original implementation \textsc{Shadow} are available, but not the actual source code, we decided to manually transform our Java subjects into C programs, so that we can check the consistency between our results and the results by \textsc{Shadow}.

The experiments were conducted on a machine running macOS 10.12.6 featuring an 2.9GHz Intel Core i5 and 16 GB of memory. We configured the symbolic executions with an unbounded depth limit and a timeout of one hour.

\subsection{Results and Analysis}\label{subsec:results-analysis}
Table \ref{table:evaluation} shows the detailed results of the performed experiments.
The first column names the corresponding class and method that were tested, together with an id which specifies each mutant.
Column \textit{type} contains the mutation operation adapted by \cite{Just2011}:  Relational Operator Replacement (ROR),  Arithmetic Operator Replacement (AOR) and Statement Deletion (STD).
Since \textsc{Major} only generated single mutants per class, we also combined them manually to get multiple changes per class.
In such cases the numbers at the end of each subject name denote the combined mutants (we named the type ``MUL'' for multiple changes).
The following columns describe the execution time in seconds, the number of visited states during the symbolic exploration, the maximum used memory in MB and the number of resulting path conditions (the number in the brackets for SPF represent the number of paths that were $\mathit{diff}$ paths) for the normal symbolic execution (SPF), for the shadow symbolic execution with our SPF extension \textsc{$Shadow_{JPF}$} (SWPF) and  for the original tool \textsc{$Shadow_{KLEE}$} (SW). The first row of the table shows the detailed execution results for the method \texttt{foo()} from Listing \ref{lst:example}.\\

\begin{table}[h]
\resizebox{\columnwidth}{!}{%
\begin{tabular}{|l|c|c|c|c|c|c|c|c|c|c|} \hline
\textbf{Subject} & \textbf{Type} &\multicolumn{2}{|c|}{\textbf{Time [$\mathbf{s}$]}} & \multicolumn{2}{|c|}{\textbf{\# States}} & \multicolumn{2}{|c|}{\textbf{Memory [$\mathbf{MB}$]}} & \multicolumn{3}{|c|}{\textbf{\# Paths (diff)}} \\ 
& & $SPF$ & $SWPF$ & $SPF$ & $SWPF$ & $SPF$ &$SWPF$& $SPF$ & $SWPF$ & $SW$ \\ \hline
Foo & - &  2 & 2 & 16 & 17 & 434 & 690 & 4 (2) & 2 & 2 \\ \hline
BankAccount.deposit\_1 & ROR & $<1$ & $<1$ & 4 & 19 & 245 & 245 & 2 (0) & 1 & 1\\ 
BankAccount.deposit\_2 & ROR & $<1$ & $<1$ & 4 & 16 & 245 & 245 & 2 (0) & 1 & 1\\ 
BankAccount.deposit\_3 & ROR & $<1$ & $<1$ & 2 & 10 & 245 & 245 & 1 (0) & 1 & 1\\ 
BankAccount.deposit\_4 & STD & $<1$ & $<1$ & 4 & 7 & 245 & 245 & 2 (0) & 1 & 1\\ 
BankAccount.deposit\_5 & AOR & $<1$ & $<1$ & 4 & 5 & 245 & 245 & 2 (0) & 0 & 0\\ 
BankAccount.deposit\_6 & AOR & $<1$ & $<1$ & 4 & 5 & 245 & 245 & 2 (0) & 0 & 0\\ 
BankAccount.deposit\_7 & AOR & $<1$ & $<1$ & 8 & 9 & 245 & 245 & 3 (1) & 1 & 1\\ 
BankAccount.deposit\_8 & STD & $<1$ & $<1$ & 4 & 5 & 245 & 245 & 2 (0) & 0 & 0 \\ \hline
BankAccount.withdraw\_1 & ROR & 1 & 1 & 6 & 19 & 245 & 245 & 3 (0) & 1 & 1\\ 
BankAccount.withdraw\_2 & ROR & <1 & <1 & 6 & 19 & 245 & 245 & 3 (0) & 1 & 1\\ 
BankAccount.withdraw\_3 & ROR & <1 & <1 & 4 & 9 & 245 & 245 & 2 (0) & 2 & 2\\ 
BankAccount.withdraw\_4 & STD & <1 & <1 & 8 & 9 & 245 & 245 & 4 (2) & 2 & 2\\ 
BankAccount.withdraw\_5 & ROR & <1 & <1 & 6 & 22 & 245 & 245 & 3 (0) & 1 & 1\\ 
BankAccount.withdraw\_6 & ROR & <1 & <1 & 6 & 22 & 245 & 245 & 3 (0) & 1 & 1\\ 
BankAccount.withdraw\_7 & ROR & <1 & <1 & 4 & 10 & 245 & 245 & 2 (0) & 1 & 1\\ 
BankAccount.withdraw\_8 & STD & <1 & <1 & 6 & 8 & 245 & 245 & 3 (0) & 0 & 0\\ 
BankAccount.withdraw\_9 & AOR & <1 & <1 & 6 & 8 & 245 & 245 & 3 (0) & 0 & 0\\ 
BankAccount.withdraw\_10 & AOR & <1 & <1 & 6 & 8 & 245 & 245 & 3 (0) & 0 & 0\\ 
BankAccount.withdraw\_11 & AOR & <1 & <1 & 8 & 12 & 245 & 245 & 4 (1) & 1 & 0\\ 
BankAccount.withdraw\_12 & STD & <1 & <1 & 6 & 8 & 245 & 245 & 3 (0) & 0 & 0\\ \hline
BankAccount.main\_1 & ROR & 2 & $<1$ & 24 & 52 & 433 & 245 & 4 (0) & 1 & 1\\ 
BankAccount.main\_2 & ROR & 2 & $<1$ & 24 & 52 & 690 & 245 & 4 (0) & 1 & 1\\ 
BankAccount.main\_3 & ROR & 2 & $<1$ & 16 & 26 & 434 & 245 & 3 (0) & 1 & 1\\ 
BankAccount.main\_4 & STD & 3 & $<1$ & 24 & 17 & 690 & 245 & 4 (1) & 1 & 1\\ 
BankAccount.main\_5 & AOR & 3 & $<1$ & 24 & 14 & 690 & 245 & 4 (0) & 0 & 0\\ 
BankAccount.main\_6 & AOR & 3 & $<1$ & 24 & 14 & 690 & 245 & 4 (0) & 0 & 0\\ 
BankAccount.main\_7 & AOR & 3 & $<1$ & 32 & 22 & 690 & 245 & 5 (0) & 0 & 0\\ 
BankAccount.main\_8 & STD & 3 & $<1$ & 24 & 14 & 690 & 245 & 4 (0) & 0 & 0\\ 
BankAccount.main\_9 & ROR & 2 & $<1$ & 24 & 58 & 434 & 245 & 4 (0) & 1 & 1\\ 
BankAccount.main\_10 & ROR & 3 & $<1$ & 24 & 58 & 690 & 245 & 4 (0) & 1 & 1\\ 
BankAccount.main\_11 & ROR & 1 & $<1$ & 16 & 26 & 433 & 245 & 3 (0) & 1 & 1\\ 
BankAccount.main\_12 & STD & 3 & $<1$ & 24 & 17 & 690 & 245 & 4 (1) & 1 & 1\\ 
BankAccount.main\_13 & ROR & 3 & $<1$ & 24 & 20 & 690 & 245 & 4 (0) & 0 & 0\\ 
BankAccount.main\_14 & ROR & 3 & $<1$ & 24 & 20 & 690 & 245 & 4 (0) & 0 & 0\\ 
BankAccount.main\_15 & ROR & 3 & $<1$ & 20 & 11 & 434 & 245 & 4 (1) & 1 & 1\\ 
BankAccount.main\_16 & STD & 3 & $<1$ & 24 & 14 & 690 & 245 & 4 (0) & 0 & 0\\ 
BankAccount.main\_17 & AOR & 2 & $<1$ & 24 & 14 & 690 & 245 & 4 (0) & 0 & 0\\ 
BankAccount.main\_18 & AOR & 2 & $<1$ & 24 & 14 & 690 & 245 & 4 (0) & 0 & 0\\ 
BankAccount.main\_19 & AOR & 3 & $<1$ & 28 & 22 & 690 & 245 & 5 (0) & 1 & 0\\ 
BankAccount.main\_20 & STD & 2 & $<1$ & 24 & 14 & 690 & 245 & 4 (0) & 0 & 0\\ 
BankAccount.main\_21 & STD & 3 & $<1$ & 24 & 17 & 690 & 245 & 4 (0) & 0 & 0\\ 
BankAccount.main\_22 & ROR & 1 & $<1$ & 8 & 15 & 309 & 245 & 2 (2) & 2 & 2\\ 
BankAccount.main\_23 & ROR & 1 & $<1$ & 8 & 15 & 309 & 245 & 2 (2) & 2 & 2\\ 
BankAccount.main\_1\_13 & MUL & 2 & $<1$ & 24 & 52 & 433 & 245 & 4 (0) & 1 & 1\\ 
BankAccount.main\_2\_22 & MUL & 1 & $<1$ & 8 & 41 & 309 & 245 & 2 (0) & 3 & 3\\ 
BankAccount.main\_15\_23 & MUL & 1 & $<1$ & 6 & 13 & 245 & 245 & 2(2) & 2 & 2\\ 
BankAccount.main\_5\_18 & MUL & 2 & $<1$ & 24 & 14 & 690 & 245 & 4 (0) & 0 & 0\\ 
BankAccount.main\_3\_23 & MUL & 1 & 1 & 8 & 15 & 309 & 245 & 2 (2) & 2 & 2\\ 
BankAccount.main\_17\_22 & MUL & 1 & 1 & 8 & 15 & 309 & 245 & 2 (2) & 2 & 2\\ 
BankAccount.main\_3\_10\_22 & MUL & $<1$ & 1 & 2 & 5 & 245 & 245 & 1 (0) & 1 & 1 \\
BankAccount.main\_5\_18\_23 & MUL & 1 & $<1$ & 8 & 15 & 309 & 245 & 2 (2) & 2 & 2\\ \hline
\end{tabular}}
\caption{Experimental results for the comparison of Symbolic PathFinder (SPF), \boldmath{\shadowjpf} (SWPF) and the original tool\boldmath{\shadowklee} (SW).}
\label{table:evaluation}
\end{table}

\noindent \textit{\textbf{RQ.1 Consistence with original\boldmath{\shadowklee}}}\\
In order to answer RQ.1 we compared the number of test cases generated by \shadowjpf~and\shadowklee~(cf. Table \ref{table:evaluation}).
In almost all cases\shadowklee~generated the same number of $\mathit{diff}$ paths.
The only differences can be observed for BankAccount.main\_11 and BankAccount.main\_19, since in both cases the $new$ version introduces a potential zero-division error and \\ \shadowklee~does not find this error (at least not with the configuration we used it).
Additionally, we also manually compared the generated path conditions, which matched for all considered subjects. \\

\noindent \textit{\textbf{RQ.2 Generating Regression Test Cases with \boldmath{\shadowjpf}\\that SPF missed}}\\
In order to answer RQ.2 we compared the number of test cases generated by pure SPF and \shadowjpf~(cf. Table \ref{table:evaluation}).
The numbers show that \shadowjpf~can reduce the number of generated test cases enormously.
All generated test cases of \shadowjpf~are real regression test cases because all of them show a divergence between the two versions. In contrast SPF generates a lot of irrelevant paths for regression testing.
This is based on the fact that SPF only can consider the information of the $old$ or the $new$ program version exclusively.
As we used the new version to run SPF, the generated path constraints are often too imprecise to trigger a real regression input, i.e., the constraint subsumes the real regression path constraint because SPF misses the crucial information from the $old$ version.
Then it depends on the value generation of the constraint solver whether the concrete test inputs hit the $\mathit{diff}$ path or not. All in all, our results show that \shadowjpf~can generate the regression test cases that were missed by SPF. \\



\section{Conclusion and Future Work}
In this work we presented our tool \shadowjpf~as an extension of the Java PathFinder project.
Our tool applies the idea of shadow symbolic execution \cite{Palikareva2016} on Java bytecode and, hence, makes shadow symbolic execution available to a large range of Java programs.
We performed preliminary experiments on 51 generated mutants and compared the results with Symbolic Path\-Finder and the original implementation\shadowklee.
They show that \shadowjpf~ can significantly reduce the number of test cases compared to SPF, and that it behaves like the original implementation in\shadowklee.
Although these results are very promising, the analyzed subjects are relatively small, and hence, we cannot generalize the results to larger and real-world examples.

In future we plan to extend our implementation and build on top of \shadowjpf~a more powerful tool for the generation of regression test cases.
This includes the full automation of the change annotation, which currently is done manually.
Additionally, we want to extend our evaluation by adding more JPF compatible classes and real-world regression bugs to our analysis.

\balance

\bibliographystyle{abbrv}
\bibliography{bibliography,regressiontesting, testaugmentation, testcasegeneration} 

\begin{thebibliography}{10}

\bibitem{Cadar2008KLEE}
C.~Cadar, D.~Dunbar, and D.~Engler.
\newblock Klee: Unassisted and automatic generation of high-coverage tests for
  complex systems programs.
\newblock In {\em Proceedings of the 8th USENIX Conference on Operating Systems
  Design and Implementation}, OSDI'08, pages 209--224, Berkeley, CA, USA, 2008.
  USENIX Association.

\bibitem{Graves2001}
T.~L. Graves, M.~J. Harrold, J.-M. Kim, A.~Porter, and G.~Rothermel.
\newblock An empirical study of regression test selection techniques.
\newblock {\em ACM Trans. Softw. Eng. Methodol.}, 10(2):184--208, Apr. 2001.

\bibitem{Gu2010}
Z.~Gu, E.~T. Barr, D.~J. Hamilton, and Z.~Su.
\newblock Has the bug really been fixed?
\newblock In {\em Proceedings of the 32Nd ACM/IEEE International Conference on
  Software Engineering - Volume 1}, ICSE '10, pages 55--64, New York, NY, USA,
  2010. ACM.

\bibitem{Harrold2001}
M.~J. Harrold, J.~A. Jones, T.~Li, D.~Liang, A.~Orso, M.~Pennings, S.~Sinha,
  S.~A. Spoon, and A.~Gujarathi.
\newblock Regression test selection for java software.
\newblock In {\em Proceedings of the 16th ACM SIGPLAN Conference on
  Object-oriented Programming, Systems, Languages, and Applications}, OOPSLA
  '01, pages 312--326, New York, NY, USA, 2001. ACM.

\bibitem{Just2011}
R.~Just, F.~Schweiggert, and G.~M. Kapfhammer.
\newblock Major: An efficient and extensible tool for mutation analysis in a
  java compiler.
\newblock In {\em Proceedings of the 2011 26th IEEE/ACM International
  Conference on Automated Software Engineering}, ASE '11, pages 612--615,
  Washington, DC, USA, 2011. IEEE Computer Society.

\bibitem{Marinescu2013}
P.~D. Marinescu and C.~Cadar.
\newblock Katch: High-coverage testing of software patches.
\newblock In {\em Proceedings of the 2013 9th Joint Meeting on Foundations of
  Software Engineering}, ESEC/FSE 2013, pages 235--245, New York, NY, USA,
  2013. ACM.

\bibitem{Palikareva2016}
H.~Palikareva, T.~Kuchta, and C.~Cadar.
\newblock Shadow of a doubt: Testing for divergences between software versions.
\newblock In {\em Proceedings of the 38th International Conference on Software
  Engineering}, ICSE '16, pages 1181--1192, New York, NY, USA, 2016. ACM.

\bibitem{Pasareanu2013}
C.~S. P{\u{a}}s{\u{a}}reanu, W.~Visser, D.~Bushnell, J.~Geldenhuys, P.~Mehlitz,
  and N.~Rungta.
\newblock Symbolic pathfinder: integrating symbolic execution with model
  checking for java bytecode analysis.
\newblock {\em Automated Software Engineering}, 20(3):391--425, 2013.

\bibitem{Visser2003}
W.~Visser, K.~Havelund, G.~Brat, S.~Park, and F.~Lerda.
\newblock Model checking programs.
\newblock {\em Automated Software Engineering}, 10(2):203--232, Apr 2003.

\bibitem{Yin2011}
Z.~Yin, D.~Yuan, Y.~Zhou, S.~Pasupathy, and L.~Bairavasundaram.
\newblock How do fixes become bugs?
\newblock In {\em Proceedings of the 19th ACM SIGSOFT Symposium and the 13th
  European Conference on Foundations of Software Engineering}, ESEC/FSE '11,
  pages 26--36, New York, NY, USA, 2011. ACM.

\end{thebibliography}

\end{document}